\documentclass[a4paper]{jpconf} 
\usepackage{graphicx}
\begin{document}
\title{Quantum Criticality at the Origin of Life}

\author{G{\'{a}}bor Vattay$^1$, Dennis Salahub$^2$,  Istv{\'{a}}n Csabai$^1$,  Ali Nassimi$^{2,3}$ \& Stuart A. Kaufmann$^{2,4}$}

\address{$^1$ Department of Physics of Complex Systems, E{\"{o}}tv{\"{o}}s University, 1117 Budapest, P{\'{a}}zm{\'{a}}ny P. s. 1/A, Hungary\\
$^2$Department of Chemistry, CMS - Centre for Molecular Simulation and IQST - Institute for Quantum Science and Technology, University of Calgary, 2500 University Drive NW, Calgary, Alberta, Canada T2N 1N4\\
$^3$ Chemical Physics Theory Group, Department of Chemistry,
University of Toronto, 80 Saint George Street,
Toronto, ON M5S 3H6, Canada\\
$^4$ Institute for Systems Biology, Seattle, WA 98109, USA}

\ead{vattay@elte.hu}

\begin{abstract}
Why life persists at the edge of chaos is a question at the very heart of evolution.
Here we show that molecules taking part in biochemical processes 
from small molecules to proteins are critical quantum mechanically.  
Electronic Hamiltonians of biomolecules are tuned exactly to the critical point of the metal-insulator transition separating the Anderson localized insulator phase from the conducting disordered metal phase. Using tools from Random Matrix Theory we confirm that the energy level statistics of these biomolecules show the universal transitional distribution of the metal-insulator critical point and the wave functions are multifractals in accordance with the theory of Anderson transitions. The findings point to the existence of a universal mechanism of charge transport in living matter. The revealed bio-conductor material is neither a metal nor an insulator but a new quantum critical material which can exist only in highly evolved systems and has unique material properties. 
\end{abstract}

Advances in the theory of complex systems over the last quarter century reinforced that living systems exist at the edge of chaos\cite{stuart1993origins,crutchfield1988computation,langton1990computation,maynard1995life} and order, poised at criticality\cite{mora2011biological}.   
Finding the detailed mechanism behind this apparent self-organized criticality\cite{lewin1993complexity,bak1988self} is still a tantalizing problem. One of the fascinating aspects of life is the highly organized molecular machinery taking care of myriads of complex processes such as DNA replication, protein synthesis, cell division and metabolism,
to mention only a few. Electric forces animating the parts require a perpetual and precise motion of charges throughout the system for perfect execution of biochemical tasks. In this paper we show that practically all biomolecules,  
from small signalling molecules to proteins taking part in biochemical electronic processes, belong to a fundamentally new class of conducting material. This is a disordered conductor where the strength of the disorder is tuned 
exactly to the metal-insulator transition point and it is consequently in a permanent critical quantum state. 

Our initial perspective is that of condensed-matter physics. The unique properties of the critical quantum state at the localization-delocalization transition point
have been described first in the  the well known Anderson\cite{anderson1958absence} model, which represents the current paradigm for understanding conduction in condensed matter. In the Anderson Hamiltonian
$H=\sum_j \epsilon_i a_i^+a_i-\sum_{<ij>}a_j^+a_i$ on a 3D lattice with random uniformly distributed on-site energies 
$\epsilon_i \in [-W,W]$, the critical level of disorder\cite{mackinnon1981one} is $W_c=16\pm0.5$. For $W<W_c$ the system is a disordered metal with extended states and for $W>W_c$ the states are
localized and the system is an insulator. It has been shown\cite{wegner1981bounds} that in the exact critical point of the Anderson transition ($W=W_c$) the electron eigenfunctions are extended, but strongly inhomogeneous multifractals\cite{castellani1986multifractal}. Similar metal-insulator transitions (MIT) exist in a wide range of physical systems at various dimensionalities\cite{evers2008anderson} including the 1D Harper-Hofstadter model\cite{hofstadter1976energy} and quantum-Hall-type transitions.

For example, the 1D Harper model\cite{harper1955general} describes the energy spectrum of an
electron in a 2D lattice, whose graphic representation
is widely known as the Hofstadter butterfly\cite{hofstadter1976energy}.
In the related Aubry-Andr{\'e} Hamiltonian\cite{aubry1980analyticity} $H=\sum_j \lambda\cos(2\pi\sigma i) a_i^+a_i-\sum_{<ij>}a_j^+a_i$, where $\sigma=(\sqrt{5}-1)/2$) is the Golden Mean Ratio, the critical value separating localized and extended phases is $\lambda_c=2$. The wave functions are also multifractals\cite{evangelou1990multifractal} there. 
Other examples include\cite{evers2008anderson} quantum-Hall-type transitions in disordered conductors and superconductors in strong magnetic fields, the spin quantum Hall effect, the thermal quantum Hall effect, Dirac fermions in random vector potentials and Bethe lattices. 


MIT-like transitions exist also in low-dimensional quantum chaos. Quantum counterparts of strongly chaotic systems\cite{bohigas1984characterization} share the properties of delocalized systems,
while integrable systems\cite{berry1977level} have localized wave functions in quantum numbers corresponding to conserved quantities. Pseudointegrable systems\cite{richens1981pseudointegrable} lie at the border of chaos and integrability, where classical
trajectories diverge only slowly (with zero Lyapunov exponent) but their dynamics is complex and
the periodic orbits proliferate in their phasespace exponentially. Quantized pseudointegrable systems also 
show all the key features of critical systems, including multifractal wavefunctions\cite{bogomolny2004structure}.

Criticality can also be observed in the energy spectrum of systems at the transition point. Random Matrix 
Theory\cite{1955,dyson1962statistical,RevModPhys.53.385,Guhr1998189} (RMT) is the main tool for the characterization of the universal statistical properties of Hamiltonians of complex systems. The distance between consecutive energy levels fluctuates 
in the spectrum. The raw distance between levels $\sigma_n=E_{n+1}-E_n$ can be normalized using the average 
separation of levels 
$\Delta(E)$ at a given energy window around $E$. The ratio $s_n=\sigma_n/\Delta(E_n)$ is called the level spacing.
Random matrix theory has certain predictions for the form of the distribution $P(s)$
of level spacings. 

It has been established\cite{PhysRevB.47.11487} that in the 3D Anderson model the localized, delocalized 
and the critical states each have a distinct level spacing distribution. These three  
distributions are believed to be universal, i.e. independent of the microscopic details of the disordered system.
In the delocalized metallic phase ($W<W_c$) the distribution coincides with the level statistics of the Gaussian Orthogonal Ensemble (GOE), which is the ensemble of real symmetric random matrices with identically distributed Gaussian 
elements. The level spacing distribution is the Wigner surmise\cite{1955}
$$P_W(s)=\frac{\pi s}{2}\exp\left(-\frac{\pi s^2}{4}\right).$$ 
In the localized insulating phase ($W>W_c$) the energy levels form a random Poisson process and the level
spacing distribution is exponential $$P_P(s)=\exp\left(-s\right).$$ 
It has been shown\cite{bohigas1984characterization,evers2008anderson,evangelou2000critical} that not only in the 3D Anderson model, but in all other examples of  Anderson-like transitions, the energy level statistics in the delocalized phase is universal and corresponds to the proper Gaussian Random Matrix Ensemble reflecting the symmetries of the system, while in the localized phase it is always random Poissonian\cite{berry1977level,evers2008anderson,evangelou2000critical}.

At the transitional point ($W=W_c$) a third kind of intermediate spectral statistics $P_T(s)$ exists, 
which is the hallmark of the critical state\cite{PhysRevB.47.11487}. In solid state models theoretical arguments\cite{aronov1994pis,aronov1995spectral} and numerical studies\cite{evangelou1994level,varga1995shape}
suggest the general form $$P_T(s)=c_1s\exp\left(-c_2s^{1+\gamma}\right),$$
where $c_1$ and $c_2$ are $\gamma$ dependent normalization constants. In the case of the 3D Anderson model\cite{varga1995shape} 
$\gamma\approx 0.2$. In the Harper model\cite{evangelou2000critical} and in
other pseudointegrable models\cite{bogomolny1999models} showing critical quantum chaos a numerical value of $\gamma\approx 0$
has been found\cite{evangelou2000critical}, which supports the semi-Poissonian distribution $$P_T(s)=P_{SP}(s)=4s\exp\left(-2s\right).$$
The semi-Poissonian distribution has also been predicted from a short range plasma model\cite{bogomolny2001short} 
of energy levels introduced in RMT.

While in physical systems the critical state can be reached only 
upon a careful tuning of the strength of the disorder, in the following we show that certain biomolecules 
are precisely  at the critical state without any external tuning.

The exact numerical solution of the Schr{\"o}dinger equation for the electronic states of large molecules such as
proteins is a prohibitive task. Various approximations have been developed, which reduce the problem to a 
one-electron problem in the effective field of the remaining electrons.
Wave functions of molecules are usually written in the form of Linear Combinations of Atomic Orbitals (LCAO)  
$\phi_i=\sum_{r}C_{ir}\chi_r$
where $\phi_i$ is a Molecular Orbital (MO) represented as the sum of atomic orbital (AO) contributions $\chi_r$.
The one-electron eigenenergies and eigenvectors then can be determined from the generalized eigenvalue equation $HC=ESC$, where $S_{rs}=\langle \chi_r\mid\chi_{s}\rangle$ and $H_{rs}=\langle\chi_r \mid\hat{H}_{eff}\mid \chi_{s}\rangle$ are the overlap and effective Hamiltonian matrices respectively. The effective Hamiltonian depends on the coefficients which makes the problem
nonlinear in $C$. This is the case in Hartree-Fock and Density Functional Theory (DFT) calculations, which then cannot be routinely carried out for proteins involving thousands of atoms.
If we restrict our interest to the localization-delocalization problem in valence electrons and 
treat the two-electron part of the Hamiltonian as in  the case of the electrons in metals,
in an average sense only, we can apply semi-empirical methods. Once the positions of the atoms are known the Extended H{\"u}ckel (EH) Molecular Orbital Method\cite{hoffmann1963extended} is quite 
successful in calculating the MOs of organic molecules. The diagonal part $H_{rr}^{(EH)}$ of the EH Hamiltonian
is given by the ionization energies of the AOs\cite{pople1970approximate}, while the off-diagonal elements
are calculated from the diagonal elements and the overlap matrix $$H^{(EH)}_{rs}=\frac{1}{2}K\left(H_{rr}+H_{ss}\right)S_{rs},$$ where the common choice for the empirical constant is $K=1.75$. This is similar in spirit to
other tight binding Hamiltonians in various models of the Anderson transition.  

Myoglobin is the first\cite{watson1969stereochemistry} and one of the best studied protein structures. It plays a central role in the oxygen storage of muscles. It consists of 153 amino acids and weighs about 18000 Daltons.
For the numerical studies we selected NMR data of a solution form\cite{osapay1994solution} (PDB ID:1MYF) from the Protein Data Bank\cite{berman2000protein} of RCSB, as it is captured in the "living" state and contains the positions of hydrogen atoms essential for the calculations. The EH calculations have been carried out by the numerical
package YAeHMOP. There are $N=6329$ valence electron AOs and the EH Hamiltonian and overlap matrices
are sufficiently large of dimension $6329 \times 6329$, which makes it possible to make a good numerical comparison
with similar calculations in solid state physics\cite{evangelou1990multifractal,varga1995shape,PhysRevB.47.11487,Guhr1998189} and quantum chaos\cite{evangelou2000critical,bogomolny2004structure,bogomolny1999models,Guhr1998189}.
L{\"o}wdin transformation of the coefficient vector $C'=S^{1/2}C$
has been applied to transform  the EH Hamiltonian to a real symmetric self-adjoint operator $H^{L}=S^{-1/2}H^{(EH)}S^{-1/2}$, which satisfies a normal eigenequation $H^{L}C^L=EC^L$.

The Highest Occupied Molecular Orbital (HOMO) and the Lowest Unoccupied Molecular Orbital (LUMO) play a key role both in electron transport and reactions of organic molecules.For the visual demonstration of the fractal nature of the eigenfunctions we show the HOMO and the LUMO of Myoglobin in Fig. 1.  Absolute values of coefficients $C^L_r$ are shown such that the index $r$ is ordered in the sequence of appearance of atoms and orbitals in the amino acid sequence of the protein. Statistical similarity of the magnified part of the wave functions to the entire function
is a visual indication of a fractal.

The multifractal analysis of the protein wave functions is based on the standard box counting procedure\cite{schreiber1991multifractal},
dividing the 1D index space of ordered AO indices along the protein sequence into $N_l\approx N/l$ boxes of size 
$l$ and determining the box probability of the wave function in the $k$th box,
$$\mu_k(l)=\sum_{n=0}^{l-1}|C_{(k-1)l+n}^L(E_i)|^2, k=1,\ldots,N_l,$$
as a suitable measaure.
If the $q$th moments of this measure are counted in all boxes and is proportional to some power $\tau(q)$ of the
box size,
$$\chi_q=\left< \sum_k \mu_k^q(l)\right>_E\sim l^{\tau(q)},$$
multifractal behaviour might be derived. For a simple monofractal $\tau(q)=(q-1)D$, where $D$ is the fractal dimension. For multifractals the $\tau(q)$ curve is nonlinear and the generalized fractal dimensions $D_q$
can be recovered $\tau(q)=(q-1)D_q=\lim_{L\rightarrow 0} \ln\chi_q/\ln l$.

We expect that for extended wave functions in the conducting phase the 
coefficients are evenly distributed around their mean, which is $\left< |C^{L}_r|^2\right>_E =(1/N)\sum_{k=1}^N | C^{L}_r(E_k) |^2=1/N,$ and are independent of the position due to the normalization of the wave functions. 
The measure then scales like $\mu_k(l)\sim l/N$ and the moments scale as $\chi_q\sim (l/N)^{(q-1)}$, yielding $D_q=D=1$
independent of $q$. For localized states in the insulating phase the coefficients are nearly zero 
except in a short interval of the size of the  localization length $\xi$. The localization length is
much smaller than the system size $\xi\ll N$, therefore at intermediate length scales $\xi \ll  l \ll N$ an 
interval of size $l$ either contains the localization interval and almost the full
probability $\mu_k(l)\approx 1$ or it is practically empty $\mu_k(l)\approx 0$. The moments do not
scale with the length $\chi_q\sim 1$ and $\tau(q)=0$ yielding $D_q=D=0$. 

\begin{figure}[!ht]
\centerline{\includegraphics[width=15cm]{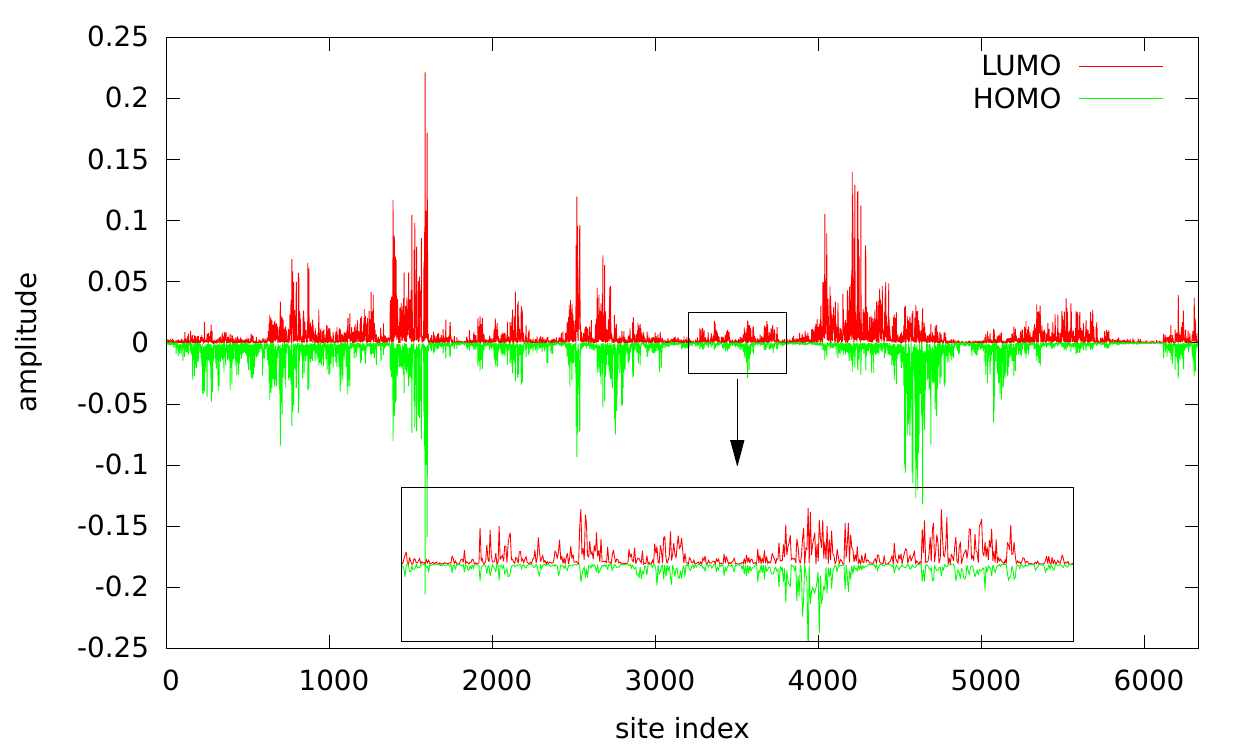}} 
\caption{{\bf The HOMO/LUMO orbitals for Myoglobin (PDB ID:1MYF) calculated with the Extended H{\"u}ckel method.} Vertical axis: the absolute value of $C^L_r$ in red for the HOMO and LUMO in green (flipped). Horizontal axis: the index sequence $r$ of AOs ordered along the amino acid sequence.  Inset: enlarged part of the box.
}
\end{figure}
The numerical values of the fractal dimension of the protein wave function can be determined more conveniently
by calculating the box probability for all possible boxes of length $l$ and performing an additional 
averaging to smooth out statistical fluctuations
$$\left<\sum_k \mu_k^q(l)\right>_E=N_l\times \frac{1}{N_l}\sum_k \left< \mu_k^q(l)\right>_E\approx \frac{N}{l}\frac{1}{N-l}\sum_{r=1}^{N-l}
\left<\left(\sum_{n=1}^{l}|C_{r+n}^L|^2\right)^q\right>_E.$$
In Fig. 2. we show the generalized fractal dimensions for Myoglobin obtained numerically. The most significant value is the correlation dimension $D_2\approx 0.5$ which is just midway between localization $D=0$ and 
delocalization $D=1$ confirming that the system is critical and the wave functions are multifractals.
We note that the same numerical value $D_2=0.5$ has been obtained also for critical quantum chaos\cite{evangelou2000critical,bogomolny2004structure}.
\begin{figure}[!ht]
\centerline{\includegraphics[width=15cm]{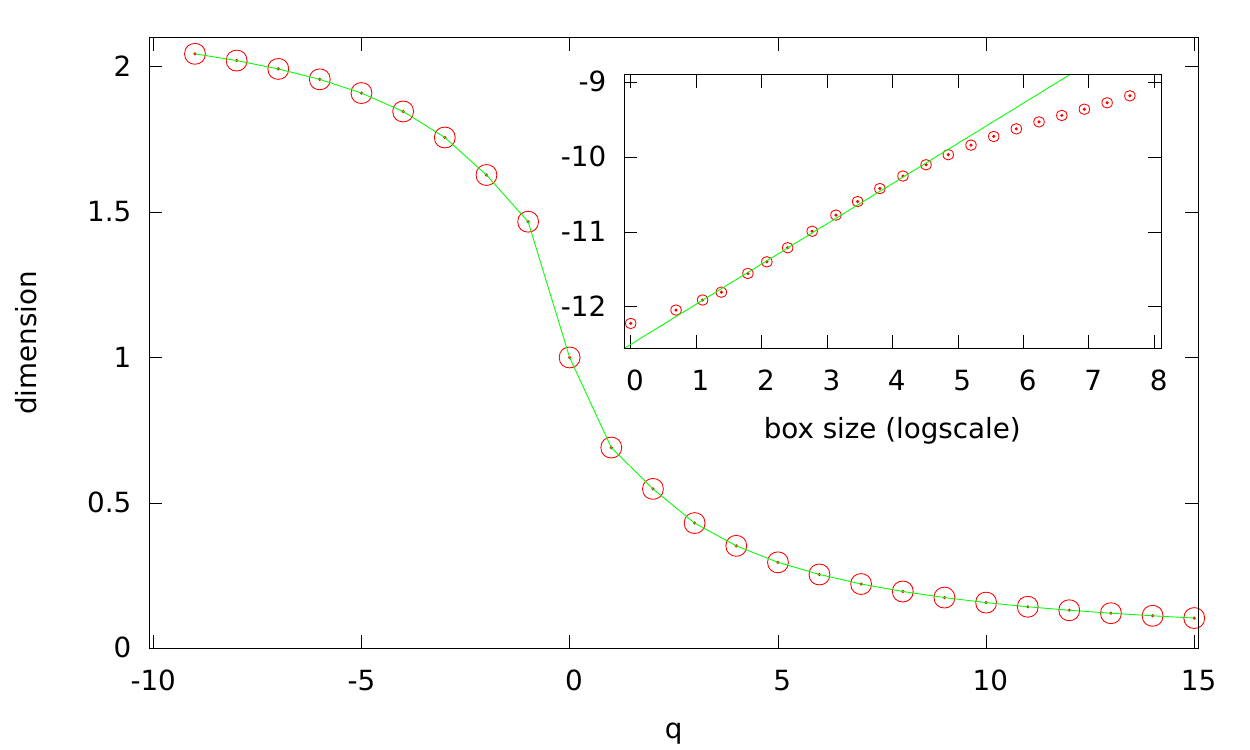}} 
\caption{{\bf Generalized fractal dimensions $D_q$ of  wave functions of the protein Myoglobin (PDB ID:1MYF) averaged
for all energies from the Extended H{\"u}ckel calculation.} Inset: Scaling of $\chi_q$ for $q=2$ as a function
of the box length $l$ on a double logarithmic plot. The green linear function is fitted to the scaling region 
and its slope yields the correlation dimension $D_2=0.5\pm 0.01$.
}
\end{figure}

Next we show that the level statistics of Myoglobin is also transitional. The energy levels computed with the 
EH method have been analyzed with a statistical method of RMT suitable for the analysis of a relatively low number
of eigenvalues. The distance between two consecutive levels $\sigma_i=E_{i+1}-E_i$
is normalized with the average of $k$ level spacings to the left and to the right $\Delta_i=\frac{1}{2k+1}\sum_{j=-k}^{j=+k} \sigma_{i+j}$ giving the unfolded level spacing $s_i=\sigma_i/\Delta_i=(2k+1)(E_{i+1}-E_i)/(E_{i+k+1}-E_{i-k})$. The choice of $k$ depends on the variability of the density of energy levels. Statistical averaging 
would require large $k$ values, while fast variation (especially singularities) in the density of energy levels
restricts our choice to low values. We have found that a choice of $k=2\ldots5$ ensures the stability of the
distribution in our examples.
Next, following standard procedures\cite{bogomolny1999models}, the cummulative spacing $I(S)=\# \{s_i<S\}/N=\int_0^SP(s)ds$ is calculated and compared to the theoretical predictions. For the Poissonian statistics $I_P(S)=1-\exp\left(-S\right)$, for the Wigner surmise $I_W(S)=1-\exp\left(-\pi S^2/4\right)$ and for the semi-Poissonian transitional
statistics $I_{SP}(S)=1-(2S+1)\exp\left(-2S\right)$. In the main part of Fig. 3. we show the cummulative level spacing for the 6328 spacings in the spectrum of Myoglobin and in the inset we show the difference to $I_{SP}(S)$.
We can see that without any parameter fitting the spacings for Myoglobin follow the critical theoretical curve with astonishing precision. Note, that no parameter fitting is involved in the procedure, the calculated spacing
distribution has a less than $3\%$ error like in the case of systems of critical quantum chaos\cite{evangelou2000critical,bogomolny2004structure}, which are purely theoretical models as opposed to our case, where the 
positions of atoms in the protein are obtained experimentally.

\begin{figure}[!ht]
\centerline{\includegraphics[width=15cm]{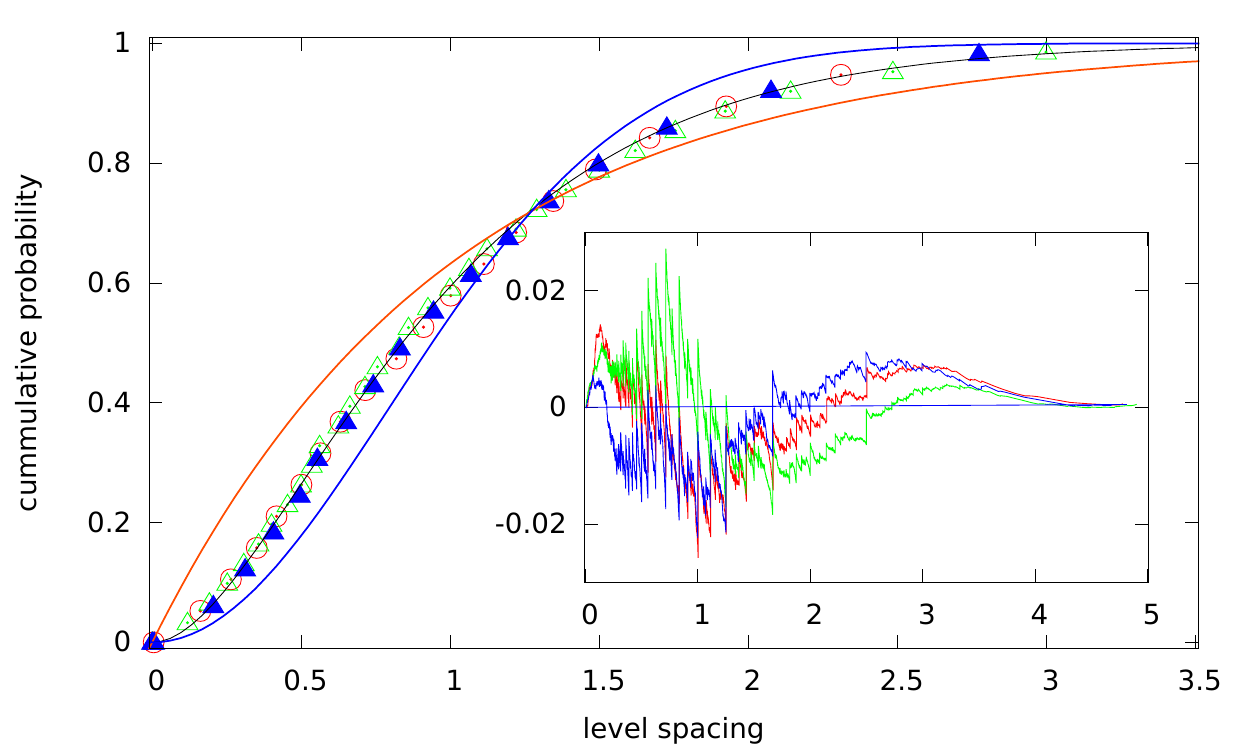}} 
\caption{{\bf Cummulative level spacing distribution} $I(S)$ for Myoglobin (blue triangles), Profilin (red circles) and Human Apolipoprotein E (green circles) are shown. Only every 300 value is shown for legibility. For comparison 
the theoretical curves $I_P(S)$ (red line), $I_W(S)$ (blue line) and $I_{SP}(S)$ (black line) are also shown.
Inset: Difference between the data curves and $I_{SP}(S)$ are shown with the colour of the data set. Each value is
plotted. The error is below $3 \%$ in probability.
}
\end{figure}
In addition to Myoglobin, we included in Fig. 3. two other proteins with known biochemical functions, selected randomly from the PDB just by size (close to 10000 valence AOs) and by the availability of the necessary 
structural data including the coordinates of Hydrogen atoms. Human profilin (PDB ID:1PFL)\cite{metzler1995refined} in solution form has $N=5232$ valence AOs. It is a ubiquitous eukaryotic protein that binds to both cytosolic actin and the phospholipid phosphatidylinositol-4,5-bisphosphate. Human apolipoprotein E (PDB ID:2L7B)\cite{chen2011topology} has $N=11980$ valence AOs.
It is one of the major determinants in lipid transport, playing a critical role in atherosclerosis and other diseases. One can see that these other randomly picked proteins are also on the critical curve with the same
precision as Myoglobin. Further analysis (not shown here) reveals essentially the same $D_2\approx 0.5$ values and the same generalized dimension $D_q$ spectra for these proteins as well. 

We should emphasize again, that finding a large tight binding Hamiltonian tuned exactly or almost exactly 
to the critical point by random chance can happen only with an astronomically low probability. So, finding just a single
protein with more than 100 amino acids having this property at random is impossible.

Next, we investigate whether criticality is restricted to certain proteins only or is it a more wide spread phenomenon.
The verification of fractality is not possible for smaller molecules as it requires a length
scale of two decades (say $l\sim 10-1000$) to fit a reliable power law to the curve $\chi_q(l)\sim l^{\tau(q)}$.
In the case of level statistics we need much less data to verify the shape of the cumulative level spacing $I(S)$.
In the RMT analysis in solid state physics and quantum chaos normally on the order of a thousand levels is
sufficient. Here we have found that using the technique developed for smaller data sets
and the usage of cumulative level statistics $I(S)$ instead of the distribution function $P(s)$, which
requires the binning of the data, jointly allow us to verify molecules with as low as $N=80$ valence AOs
reliably.  In Fig. 4. we show a collection of level statistics coming from organic molecules of various size.
The 3D structures of small molecules were taken from PubChem\cite{bolton2008pubchem} and the energies
are calculated with the EH method (YAeHMOP).
\begin{figure}[!ht]
\centerline{\includegraphics[width=15cm]{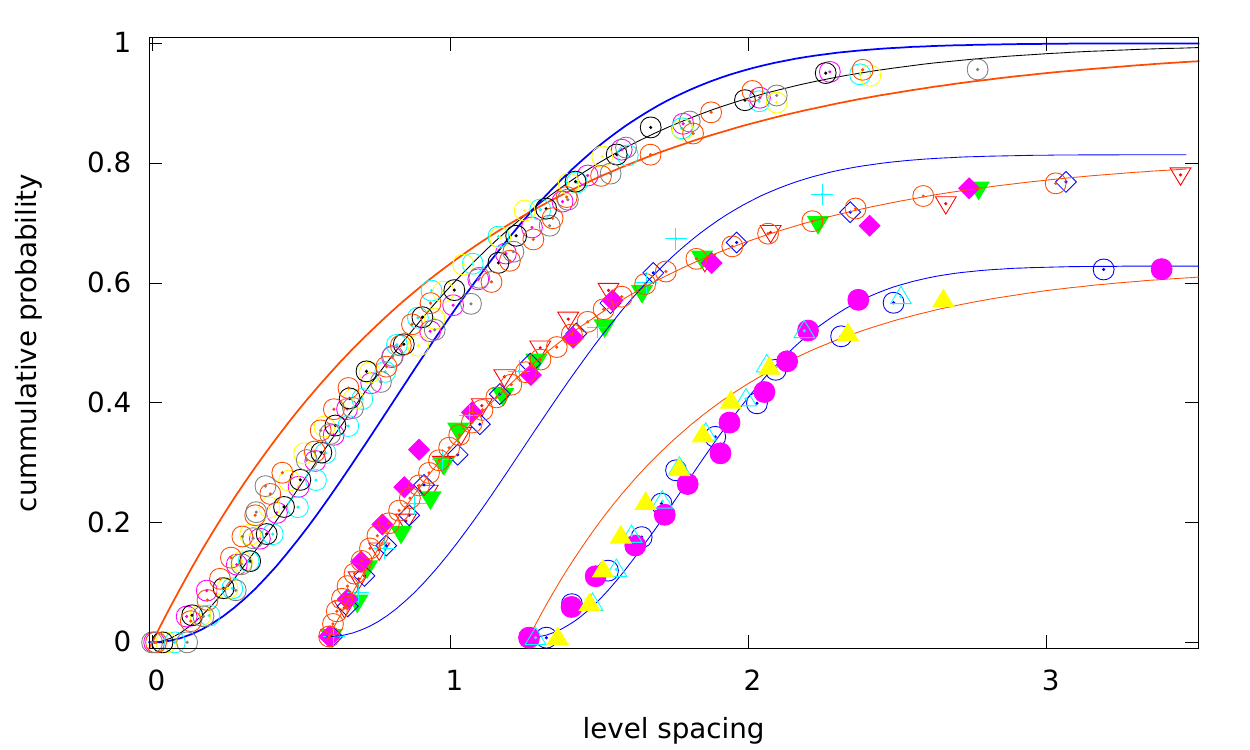}} 
\caption{{\bf Cummulative level spacing distribution for various molecules. }
(Only 10-15 data points for each
molecule is shown for legibility.) The main part: Molecules with critical cummulative level statistics $I_{SP}(S)=1-(2S+1)e^{-2S}$. The molecules shown are Vitamin B12, Vitamin D3, Linoleic Acid, Primary fluorescent chlorophyll catabolite, Sucrose and Leucine. Middle part: Molecules with Poissonian cummulative level statistics
$I_P(S)=1-e^{-S}$ (Notice, that the inset is zoomed and the main axis scales don't apply). The molecules shown are Dextrin, Silk, Octadecane, Gefarnate and a 21 base pair DNA sequence (NDB ID: 2JYK). Small part: Molecules with Wigner cummulative level statistics
$I_W(S)=1-e^{-\pi S^2/4}$ (Notice, that the inset is zoomed and the main axis scales don't apply). The molecules shown are Testosterone, Progesterone, Dibenzo(a,e)pyrene 
and Aristolochic Acid. 
}
\end{figure}

Our first and most striking observation is that each molecule investigated from the biological domain
belongs to one of the "clean" categories $I_P, I_W$ or $I_{SP}$. The reason is
by no means obvious. In RMT\cite{Guhr1998189} and especially in quantum chaos\cite{berry1984semiclassical} 
intermediate distributions between the classes occur in finite systems. It seems that the Hamiltonians
of these biomolecules are not random, they are tuned firmly to one of the classes.

Our first example is Silk (PDB ID:1SLK)\cite{fossey1991conformational}, a protein which serves as a structural material and does not take an active part in biochemical processes. In Fig. 4. we can see that it has Poissonian level statistics and it belongs to the localized class. This is in line with the fact that silk is 
a very good  insulator. This example confirms that criticality is not the property of individual amino acids. 
Amino acid sequences can produce not just critical materials but insulators as well. 

Other examples for structural biomaterials in Fig. 4. are Dextrin (CID 62698), which is a gum like substance and  Octadecane (CID 11635) which is an alkane hydrocarbon found in mineral oil and Gefarnate (CID 5282182)
a water insoluble terpene fatty acid. They all show Poissonian level statistics and are good insulators.

It is less obvious why DNA belongs to this category. In Fig. 4. we show level statistics for a 21 basis pair
DNA sequence (NDB ID:2JYK)\cite{masliah2008identification} which is clearly Poissonian.

The conductivity properties of DNA or RNA sequences are highly debated, however there seems to be a consensus that native DNA is a wide band gap semiconductor, practically an 
insulator\cite{mallajosyula2010toward}. 

We picked a few representatives from essential classes of biomolecules. From each class we used
the molecule with the largest number of AOs possible among all possible molecules having complete 
3D crystallographic data in PubChem. They all show criticality and semi-Poissonian statistics.
The list includes Linoleic acid (CID 5280450), Primary fluorescent chlorophyll catabolite (CID 54740347),
Sucrose (CID 5988), Vitamin D3 (CID 25245915), Vitamin B12 (CID 16212801) and
the largest amino acid Leucine (CID 6106). We investigated a dozen more biomolecules of
lesser size. For all practical purposes they all showed semi-Poissonian statistics, but the
statistics had larger errors due to the small number of levels. This list includes Nicotine,
Adenosine, Caffeine, Amphetamine, Benzoanthracene, Chlorpozamine, Glucose, Fatty acids omega 3 and omega 6, Picrotin, Picrotoxin, Theophylline, Thiactin, Xanthine. We also carried out the analysis for the 20 amino
acids coded by the universal genetic code. They are generally too small for the analysis of the level 
statistics to the level shown in Fig. 4. but some of them can already be classified by inspection. 
Based on this it is likely that Arginine, Cysteine, Sele Histidine, Isoleucine, Leucine, Methionine, Phenylalanine, Proline, Serine, Threonine and Tryptophan are critical and Alanine, Asparagine, Aspartic-acid, Glutamic-acid, Lysine, Tyrosine and Valine show Poissonian level statistics. For Glutamine and Glycine the results are inconclusive.

We can also find molecules which belong to the good conductor class with Wigner level spacing
statistics. We could find this only in polycyclic molecules with delocalized wave functions spreading through 
the entire molecule. In Fig. 4. we show Testosterone (CID 6013), Progesterone (CID 5994), Dibenzo(a,e)pyrene (CID 9126) and Aristolochicacid (CID 2236). Among biomolecules we could find only Steroids
in this class, the rest of such molecules were involved in combustion such as polycyclic aromatic hydrocarbons 
and were toxic or carcinogenic.

We can summarize these findings as follows: Most of the molecules taking part actively in
biochemical processes are tuned exactly to the transition point and are critical conductors. 
There is only the special class of polycyclic molecules with closely packed aromatic rings which 
show metallic behaviour and delocalization and the class of "structural materials" which play a
role in the mechanical stiffness of biological systems. Individual amino acids can be Poissonian
or critical but they seem to form proteins and polypeptides which are also either Poissonian or
critical.

These findings suggest an entirely new and universal mechanism of conductance in biology very different
from the one used in electrical circuits. In metallic conductors charges float
due to voltage differences. The electrical field accelerates electrons while scattering on impurities 
dissipates their energy fixing a constant average propagation velocity. In biological systems
we seldom see examples for this. A more likely scenario is that a charge entering a critical
conductor biomolecule will be under the joint influence of the quantum Hamiltonian and the excessive decoherence
caused by the environment\cite{de2012quantum}. Such conductance mechanism has been found for the 
excitons in light harvesting systems\cite{engel2007evidence} and it is currently in the focus of 
research in Quantum Biology\cite{huelga2014quantum}. In these systems Environment-assisted Quantum Transport\cite{rebentrost2009environment} (ENAQT) is dominant and facilitates the fast quantum spreading of 
excitations over the system. We think that this mechanism is more universal in biological systems and
charges in biological conductors are also subjects of this transport mechanism.
Recently we have shown\cite{vattay2014quantum} that ENAQT is the most effective
at the critical point of the localization-delocalization transition and the excitonic Hamiltonians
of light harvesting systems are also at or near the critical point\cite{vattay2013evolutionary}.
In the localized regime transport is hindered by strong quantum effects, while in delocalized 
systems decoherence destroys quantum propagation and the anti-Zeno effect\cite{rebentrost2009environment} 
slows down diffusion. At the mobility edge the existence of extended multifractal wave functions 
throughout the system ensure end-to-end transport while coherence decays only algebraically\cite{vattay2014quantum} ensuring a longer coherence time and supressing the anti-Zeno effect.

Our results also suggest that quantum transport played a distinguished role in evolution and 
selection. Both the number of known small molecules and proteins is about $10^8$ and the number of chemically feasible small ($<500 Da$) organic compounds is astronomical, estimated\cite{reymond2012exploring} to be
$10^{60}$. The number of proteins grows exponentially with the 
number $n$ of  amino acids as $\sim 20^n$, and the largest known has about $n\approx 26000$.
This shows that chemical and biological evolution selected only a tiny fraction $p\sim 10^{-50}$
of possible small biomolecules and even less for proteins. As the probability of finding a critical molecule or protein 
by random chance is also astronomically low, the large number of critical molecules and proteins found by quasi random browsing of major databases just by size and availability of the 3D crystallographic data suggests that 
criticality of the quantum Hamiltonian is prevailing in the evolutionary selection of biomolecules.
Besides, the fact that some proteins are natural critical conductors may open up new avenues in materials 
science as well.


\section*{References}
\bibliography{DICE2014}

\providecommand{\newblock}{}
\begin{thebibliography}{10}
\expandafter\ifx\csname url\endcsname\relax
  \def\url#1{{\tt #1}}\fi
\expandafter\ifx\csname urlprefix\endcsname\relax\def\urlprefix{URL }\fi
\providecommand{\eprint}[2][]{\url{#2}}

\bibitem{stuart1993origins}
Kauffman S~A 1993 {\em The origins of order: Self-organization and selection in
  evolution\/} (Oxford university press)

\bibitem{crutchfield1988computation}
Crutchfield J~P and Young K 1988 {\em The Santa Fe Institute, Westview\/}
  (Citeseer)

\bibitem{langton1990computation}
Langton C~G 1990 {\em Physica D: Nonlinear Phenomena\/} {\bf 42} 12--37

\bibitem{maynard1995life}
Maynard~Smith J 1995 {\em The New York review of books\/} {\bf 42} 28--30

\bibitem{mora2011biological}
Mora T and Bialek W 2011 {\em Journal of Statistical Physics\/} {\bf 144}
  268--302

\bibitem{lewin1993complexity}
Lewin R and Bak P 1993 {\em American Journal of Physics\/} {\bf 61} 764--765

\bibitem{bak1988self}
Bak P, Tang C and Wiesenfeld K 1988 {\em Physical review A\/} {\bf 38} 364

\bibitem{anderson1958absence}
Anderson P~W 1958 {\em Physical review\/} {\bf 109} 1492

\bibitem{mackinnon1981one}
MacKinnon A and Kramer B 1981 {\em Physical Review Letters\/} {\bf 47} 1546

\bibitem{wegner1981bounds}
Wegner F 1981 {\em Zeitschrift f{\"u}r Physik B Condensed Matter\/} {\bf 44}
  9--15

\bibitem{castellani1986multifractal}
Castellani C and Peliti L 1986 {\em Journal of physics A: mathematical and
  general\/} {\bf 19} L429

\bibitem{evers2008anderson}
Evers F and Mirlin A~D 2008 {\em Reviews of Modern Physics\/} {\bf 80} 1355

\bibitem{hofstadter1976energy}
Hofstadter D~R 1976 {\em Physical review B\/} {\bf 14} 2239

\bibitem{harper1955general}
Harper P 1955 {\em Proceedings of the Physical Society. Section A\/} {\bf 68}
  879

\bibitem{aubry1980analyticity}
Aubry S and Andr{\'e} G 1980 {\em Ann. Israel Phys. Soc\/} {\bf 3} 37

\bibitem{evangelou1990multifractal}
Evangelou S 1990 {\em Journal of Physics A: Mathematical and General\/} {\bf
  23} L317

\bibitem{bohigas1984characterization}
Bohigas O, Giannoni M~J and Schmit C 1984 {\em Physical Review Letters\/} {\bf
  52} 1

\bibitem{berry1977level}
Berry M~V and Tabor M 1977 {\em Proceedings of the Royal Society of London. A.
  Mathematical and Physical Sciences\/} {\bf 356} 375--394

\bibitem{richens1981pseudointegrable}
Richens P and Berry M 1981 {\em Physica D: Nonlinear Phenomena\/} {\bf 2}
  495--512

\bibitem{bogomolny2004structure}
Bogomolny E and Schmit C 2004 {\em Physical review letters\/} {\bf 92} 244102

\bibitem{1955}
Wigner E~P 1955 {\em Annals of Mathematics\/} {\bf 62} pp. 548--564 ISSN
  0003486X \urlprefix\url{http://www.jstor.org/stable/1970079}

\bibitem{dyson1962statistical}
Dyson F~J 1962 {\em Journal of Mathematical Physics\/} {\bf 3} 140--156

\bibitem{RevModPhys.53.385}
Brody T~A, Flores J, French J~B, Mello P~A, Pandey A and Wong S~S~M 1981 {\em
  Rev. Mod. Phys.\/} {\bf 53}(3) 385--479
  \urlprefix\url{http://link.aps.org/doi/10.1103/RevModPhys.53.385}

\bibitem{Guhr1998189}
Guhr T, Müller–Groeling A and Weidenmüller H~A 1998 {\em Physics Reports\/}
  {\bf 299} 189 -- 425 ISSN 0370-1573
  \urlprefix\url{http://www.sciencedirect.com/science/article/pii/S0370157397000884}

\bibitem{PhysRevB.47.11487}
Shklovskii B~I, Shapiro B, Sears B~R, Lambrianides P and Shore H~B 1993 {\em
  Phys. Rev. B\/} {\bf 47}(17) 11487--11490
  \urlprefix\url{http://link.aps.org/doi/10.1103/PhysRevB.47.11487}

\bibitem{evangelou2000critical}
Evangelou S and Pichard J~L 2000 {\em Physical review letters\/} {\bf 84} 1643

\bibitem{aronov1994pis}
Aronov A, Kravtsov V and Lerner I 1994 {\em JETP Lett\/} {\bf 59} 40

\bibitem{aronov1995spectral}
Aronov A~G, Kravtsov V~E and Lerner I~V 1995 {\em Physical review letters\/}
  {\bf 74} 1174

\bibitem{evangelou1994level}
Evangelou S 1994 {\em Physical Review B\/} {\bf 49} 16805

\bibitem{varga1995shape}
Varga I, Hofstetter E, Schreiber M and Pipek J 1995 {\em Physical Review B\/}
  {\bf 52} 7783

\bibitem{bogomolny1999models}
Bogomolny E, Gerland U and Schmit C 1999 {\em Physical Review E\/} {\bf 59}
  R1315

\bibitem{bogomolny2001short}
Bogomolny E, Gerland U and Schmit C 2001 {\em The European Physical Journal
  B-Condensed Matter and Complex Systems\/} {\bf 19} 121--132

\bibitem{hoffmann1963extended}
Hoffmann R 1963 {\em The Journal of Chemical Physics\/} {\bf 39} 1397--1412

\bibitem{pople1970approximate}
Pople J~A and Beveridge D~L 1970 {\em Approximate molecular orbital theory\/}
  vol~30 (McGraw-Hill New York)

\bibitem{watson1969stereochemistry}
Watson H~C 1969 {\em Prog. Stereochem\/} {\bf 4} 299--333

\bibitem{osapay1994solution}
{\"O}sapay K, Theriault Y, Wright P~E and Case D~A 1994 {\em Journal of
  molecular biology\/} {\bf 244} 183--197

\bibitem{berman2000protein}
Berman H~M, Westbrook J, Feng Z, Gilliland G, Bhat T, Weissig H, Shindyalov I~N
  and Bourne P~E 2000 {\em Nucleic acids research\/} {\bf 28} 235--242

\bibitem{schreiber1991multifractal}
Schreiber M and Grussbach H 1991 {\em Physical review letters\/} {\bf 67} 607

\bibitem{metzler1995refined}
Metzler W~J, Farmer B~T, Constantine K~L, Friedrichs M~S, Mueller L and Lavoie
  T 1995 {\em Protein Science\/} {\bf 4} 450--459

\bibitem{chen2011topology}
Chen J, Li Q and Wang J 2011 {\em Proceedings of the National Academy of
  Sciences\/} {\bf 108} 14813--14818

\bibitem{bolton2008pubchem}
Bolton E~E, Wang Y, Thiessen P~A and Bryant S~H 2008 {\em Annual reports in
  computational chemistry\/} {\bf 4} 217--241

\bibitem{berry1984semiclassical}
Berry M~V and Robnik M 1984 {\em Journal of Physics A: Mathematical and
  General\/} {\bf 17} 2413

\bibitem{fossey1991conformational}
Fossey S~A, N{\'e}methy G, Gibson K~D and Scheraga H~A 1991 {\em Biopolymers\/}
  {\bf 31} 1529--1541

\bibitem{masliah2008identification}
Masliah G, Ren{\'e} B, Zargarian L, Fermandjian S and Mauffret O 2008 {\em
  Journal of molecular biology\/} {\bf 381} 692--706

\bibitem{mallajosyula2010toward}
Mallajosyula S~S and Pati S~K 2010 {\em The Journal of Physical Chemistry
  Letters\/} {\bf 1} 1881--1894

\bibitem{de2012quantum}
de~la Lande A, Babcock N~S, {\v{R}}ez{\'a}{\v{c}} J, L{\'e}vy B, Sanders B~C
  and Salahub D~R 2012 {\em Physical Chemistry Chemical Physics\/} {\bf 14}
  5902--5918

\bibitem{engel2007evidence}
Engel G~S, Calhoun T~R, Read E~L, Ahn T~K, Man{\v{c}}al T, Cheng Y~C,
  Blankenship R~E and Fleming G~R 2007 {\em Nature\/} {\bf 446} 782--786

\bibitem{huelga2014quantum}
Huelga S~F and Plenio M~B 2014 {\em Nature Physics\/}

\bibitem{rebentrost2009environment}
Rebentrost P, Mohseni M, Kassal I, Lloyd S and Aspuru-Guzik A 2009 {\em New
  Journal of Physics\/} {\bf 11} 033003

\bibitem{vattay2014quantum}
Vattay G, Kauffman S and Niiranen S 2014 {\em PloS one\/} {\bf 9} e89017

\bibitem{vattay2013evolutionary}
Vattay G and Kauffman S~A 2013 {\em arXiv preprint arXiv:1311.4688\/}

\bibitem{reymond2012exploring}
Reymond J~L and Awale M 2012 {\em ACS chemical neuroscience\/} {\bf 3} 649--657

\end{thebibliography}
\bibliographystyle{iopart-num}

\end{document}